\documentclass[11pt,draftclsnofoot,journal,onecolumn]{IEEEtran}

\usepackage{cite}
\usepackage{url}

\usepackage{graphicx}
\usepackage{amssymb}
\usepackage{amsmath}
\usepackage{color}
\usepackage{alltt}
\usepackage{tikz}
\usepackage{pgfplots}
\usepackage{pgf-umlsd}
\usepackage{cprotect}
\usepackage{multirow}
\usepackage{algorithm}
\usepackage[noend]{algpseudocode}
\usepackage{bytefield}
\usepackage{bigfoot}
\usepackage{paralist}
\usepackage{subfigure}
\usepackage{threeparttable}
\usepackage{microtype}
\usepackage{flushend}

\usepackage{cleveref}

\makeatletter
\let\MYcaption\@makecaption
\makeatother

\makeatletter
\let\@makecaption\MYcaption
\makeatother

\usetikzlibrary{arrows,shadows,positioning,automata, calc, shapes.misc}

\tikzset{cross/.style={cross out, draw=black, minimum size=2*(#1-\pgflinewidth), inner sep=0pt, outer sep=0pt},cross/.default={3pt}}

\IEEEoverridecommandlockouts

\newtheorem{lemma}{Lemma}
\newtheorem{proposition}{Proposition}

\newtheorem{evidence}{Evidence}

\renewcommand{\IEEEQED}{\IEEEQEDopen}

\begin{document}
\title{Wisent: Robust Downstream Communication and Storage for Computational RFIDs}
\author{\IEEEauthorblockN{Jethro Tan\IEEEauthorrefmark{1}, Przemys{\l}aw Pawe{\l}czak\IEEEauthorrefmark{1}, Aaron Parks\IEEEauthorrefmark{2}, and Joshua R. Smith\IEEEauthorrefmark{2}}

\IEEEauthorblockA{\IEEEauthorrefmark{1}Delft University of Technology, Mekelweg\,4, 2628\,CD Delft, the Netherlands \\ Email: \{j.e.t.tan, p.pawelczak\}@tudelft.nl}

\IEEEauthorblockA{\IEEEauthorrefmark{2}University of Washington, Seattle, WA 98195-2350, USA\\ Email: anparks@uw.edu, jrs@cs.uw.edu}
\thanks{\copyright~2016 IEEE. Personal use of this material is permitted. Permission from IEEE must be obtained for all other uses, in any current or future media, including reprinting/republishing this material for advertising or promotional purposes,creating new collective works, for resale or redistribution to servers or lists, or reuse of any copyrighted component of this work in other works.}
\thanks{Supported by the Dutch Technology Foundation STW grant 12491 and in part by US NSF grant CNS 1305072, CNS 1407583 and EEC 1028725, Intel Corporation, and the Google Faculty Research Awards program.}}

\maketitle

\begin{abstract}
Computational RFID (CRFID) devices are emerging platforms that can enable perennial computation and sensing by eliminating the need for batteries.
Although much research has been devoted to improving upstream (CRFID to RFID reader) communication rates, the opposite direction has so far been neglected, presumably due to the difficulty of guaranteeing fast and error-free transfer amidst frequent power interruptions of CRFID.
With growing interest in the market where CRFIDs are forever-embedded in many structures, it is necessary for this void to be filled.
Therefore, we propose Wisent---a robust downstream communication protocol for CRFIDs that operates on top of the legacy UHF RFID communication protocol: EPC C1G2.
The novelty of Wisent is its ability to adaptively change the frame length sent by the reader, based on the length throttling mechanism, to minimize the transfer times at varying channel conditions.
We present an implementation of Wisent for the WISP\,5 and an off-the-shelf RFID reader.
Our experiments show that Wisent allows transfer up to 16 times faster than a baseline, non-adaptive shortest frame case, i.e. single word length, at sub-meter distance.
As a case study, we show how Wisent enables wireless CRFID reprogramming, demonstrating the world's first wirelessly reprogrammable (software defined) CRFID.
\end{abstract}


\section{Introduction}
\label{sec:introduction}

\bstctlcite{BSTcontrol}Computational RFIDs (CRFIDs) and other wireless, transiently-powered computing devices are emerging platforms that enable sensing, computation and communication without batteries~\cite{gollakota:2014:computer}.
CRFIDs have been proposed for use in scenarios such as structural health monitoring, in which a device may be embedded in a reinforced concrete structure~\cite{jiang:2011:optimum}, or implanted scenarios involving tasks such as blood glucose monitoring~\cite{xiao:2015:implantable} or pacemaker control~\cite{halperin:sap:2008}, where access to the device requires an expensive and/or risky surgery.
In these deeply embedded applications, maintenance of the device is difficult or impossible, making battery-free operation attractive as the battery maintenance requirement is avoided entirely.

However, as the complexity of use cases for CRFIDs grow, there is another emerging maintenance requirement: the need to patch or replace the firmware of the device, or to alter application parameters including the RFID radio layer controls.
In current CRFIDs, maintenance of firmware (due to e.g. errors) requires a physical connection to CRFID, nulling the main benefit of battery-free operation.

\begin{figure}
\centering
\includegraphics[width=0.65\columnwidth]{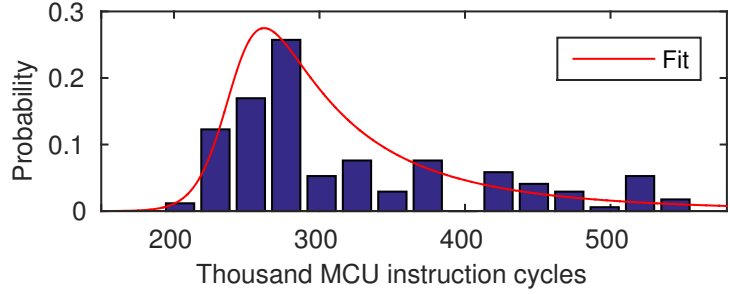}
\vspace{-2.5mm}
\caption{Example distribution of CRFID (WISP\,5~\cite{wisp5}) available harvested energy between periods of state loss (experiment details:~\cite{wisent:source_code:2016}). Energy is mapped into the number of instruction cycles WISP\,5 can achieve per burst with its default hardware configuration. The frequent state loss makes reliable bulk data transfer hard.}
\label{fig:powercycles}
\vspace{-2.5mm}
\end{figure}

Existing CRFIDs have no means for reliable high-rate downstream (i.e. RFID reader to CRFID) wireless communication and storage. 
This makes downstream protocols for CRFIDs a potent area of study.
Current CRFIDs use UHF RFID standards, such as EPC C1G2~\cite{epc2015}, designed to support inventory management applications with minimal computational and data transfer requirements~\cite{smith2013,zhang2011}.
Any UHF RFID network is built around an interrogator, which provides power to tags in the vicinity, and which can both send/receive data to/from those tags.
The uplink from tag to reader is accomplished through backscatter communication, in which the tag modulates its reflection of the reader's carrier signal in order to communicate---with orders of magnitude less power than a conventional radio. 
The reader is a wall socket-powered device, while the tag is a highly energy-constrained energy-harvesting battery-free platform, resulting in frequent power supply breaks, see~\Cref{fig:powercycles}. 

\subsection{Problem Statement}

In CRFIDs, downstream transfer functionality would enable \begin{inparaenum}[(i)]\item the processing of data sent wirelessly, \item CRFID to CRFID data transmission, and most importantly \item wireless reconfiguration, including reconfiguration of the communication protocol itself (e.g. demodulation, decoding)\end{inparaenum}.
In this work, we therefore pose the following research question: \emph{How to enable robust and fast downstream communication for CRFIDs?}

There are two challenges in enabling CRFID downstream.

\textbf{Challenge 1:} Existing CRFIDs rely on the suboptimal downstream communication capability of the EPC C1G2 standard.
In fact, EPC C1G2 does not have built-in support for \emph{fast} transfers of large portions of data from the reader to the tag.
Building a replacement for reader to CRFID tag communication protocol would indeed improve communication capabilities and performance, however, doing so would reduce compatibility with existing EPC C1G2-compliant RFID systems, which represent a significant amount of existing infrastructure.
Therefore, designing a protocol extension \emph{on top of} EPC C1G2 is the natural path (although a difficult one) towards a capable yet practical downstream CRFID protocol.

\textbf{Challenge 2:} Transient power availability in CRFIDs means that tags will often lose power, and in turn processor state~\cite{ransford2011} (again see~\Cref{fig:powercycles}).
Introduction of FRAM in latest CRFID release, i.e. WISP\,5~\cite{wisp5}, alleviated slow read/write operations of non-volatile flash memory of previous CRFID releases~\cite{ransford2010}. The question remains whether FRAM will suffice for repetitive data storage operations in CRFID.

\subsection{Contributions}

While the current focus of CRFID systems lies in improving \emph{upstream} communication, see e.g.~\cite{naderiparizi2015,zheng2014,gummeson2012}, due to the mentioned challenges, to the best of our knowledge, there has surprisingly been no work focusing on \emph{downstream} communication, where CRFIDs are on the receiving end of large data transfers.
To fill this void our contributions are:

\textbf{Contribution 1:} We leverage EPC C1G2 functionality to implement and evaluate multi-word downstream data transfer, i.e. longer than the de facto limit of a 16-bit word (2\,bytes).
Experimentally we show threefold improvement of raw downstream throughput.
We also show that the error rate of multi-word messages transferred by the reader to the CRFID reduces throughput instantly to zero at long reader to CRFID distances.

\textbf{Contribution 2:} We design, implement and experimentally evaluate Wisent\footnote{Wisent, i.e. the European Bison, is a wordplay on '\textbf{Wi}relessly \textbf{sent}'.}---a downstream-oriented protocol for CRFIDs. 
Wisent allows for an adaptation of the length of the reader message size based on the received message rate at the CRFID to keep the transfer speed high, while minimizing the data transfer errors.
Furthermore, utilizing FRAM's capability of random-access instantaneous read/write operations without memory block erasure available in WISP\,5~\cite{wisp5}, we enhance Wisent with a mechanism for fast storage and verification of large portions of data.

\textbf{Contribution 3:} We implement and demonstrate the world's first fully wirelessly-reprogrammable CRFID, enabling software-defined CRFID with truly reprogrammable radio stack.

The remainder of the paper is organized as follows.
Related work is described in~\Cref{sec:related}, while~\Cref{sec:design} introduces the CRFID downstream protocol preliminaries.
\Cref{sec:implementation} outlines Wisent design and its implementation. Experimental results and evaluation of CRFID downstream transfer improvements through the EPC C1G2 \verb|BlockWrite| implementation, Wisent itself, and the use of Wisent in wireless reprogramming scenarios are presented in~\Cref{sec:evaluation}.
Future work is discussed in~\Cref{sec:futurework}, while the paper is concluded in~\Cref{sec:conclusion}.

Finally, we remark that in the spirit of the results reproducibility, source code for Wisent, measurement data and log files are available upon request or via~\cite{wisent:source_code:2016}.


\section{Related Work}
\label{sec:related}

\subsection{Data Transfer on CRFID}
\label{sec:data_transfer_crfid}

As we noted earlier, no existing work, to the best of our knowledge, targets fast and reliable reader-to-CRFID communication.
On the other hand various CRFID-related works discuss data schemes for upstream (tag-to-reader) data transfers.
Examples of such work include HARMONY~\cite{zheng2014}, a data transfer scheme implemented on the WISP 4.1 with the purpose of reading bulk data, and Flit~\cite{gummeson2012}, a coordinated bulk transfer protocol implemented on the Intel WISP.
A recent work~\cite{naderiparizi2015} considers the image data transfer from the camera-enabled WISP to an RFID host.
For the interested reader we refer to~\cite{buettner2008,mohaisen2008}, where system-level evaluation of link layer performance of EPC C1G2 has been investigated.
We also refer to BLINK~\cite{zhang2012}: the only known alternative to EPC C1G2 in the context of CRFID, and to~\cite{buettner:rfid:2011}, where a modified EPC C1G2 was used to speed up tag access operations.

\subsection{Wireless Reprogramming of Communication Platforms}

The concept of wireless reprogramming and over-the-air (OTA) firmware update mechanisms is well researched in non-CRFID fields, namely in cellular systems and wireless sensor networks (WSNs). 
For example, a software redistribution architecture for mobile cellular terminals was discussed first in~\cite{dillinger:commag:2002}, while code distribution architecture for WSNs has been discussed in~\cite{reijers:wsna:2003}.
However, wireless reprogramming is completely new to the CRFID field and the only two relevant works we are aware of are Bootie~\cite{ransford2010} and FirmSwitch~\cite{yang2015}.
Bootie describes a proof-of-concept bootloader for CRFIDs and a preliminary design of a firmware update protocol that allows Bootie to accept and install firmware updates wirelessly.
However, its wireless protocol proposal neglected error handling and bookkeeping, has not been implemented or tested and no results on its performance exist thus far.
FirmSwitch introduced and implemented a wireless firmware switching mechanism for CRFID.
Unfortunately FirmSwitch does not deliver the most significant benefit of the reprogramming, as only pre-installed programs can be selected by its CRFID bootloader. Neither OTA firmware updates nor fast/reliable transmission protocols were proposed therein.


\section{Downstream CRFID Protocol: Preliminaries}
\label{sec:design}

\subsection{Host-to-CRFID Communication: System Overview}
\label{sec:sysoverview}

Three fundamental components of a typical CRFID system are (i) a host machine---any device that is able to communicate with the RFID reader over any popular physical interfaces such as Ethernet or Wi-Fi, (ii) an RFID reader (connected to an antenna) and (iii) a CRFID tag, see~\Cref{fig:systemsetup}.
All communications from a host to a tag are first sent to the reader. 
In the context of the EPC C1G2~\cite{epc2015} standard, the Low Level Reader Protocol (LLRP)~\cite{epc2010} specifies a network interface for such communication.
The reader then issues EPC C1G2 commands corresponding to the LLRP messages it received from the host.
Naturally, EPC C1G2 and LLRP communication primitives enforce limitations that have to be taken into account in any CRFID protocol design.
We define them below.

\tikzstyle{a} = [rectangle, draw, node distance=8mm, fill=white!20, text width=12mm, text centered, rounded corners, minimum height=2em, thick]
\tikzstyle{b} = [rectangle, draw, node distance=8mm, fill=white!20, text width=15mm, text centered, rounded corners, minimum height=4em, thick]
\tikzstyle{l} = [draw, -latex',thick]

\begin{figure}
\centering
	\subfigure[CRFID hardware]{\includegraphics[width=0.55\columnwidth]{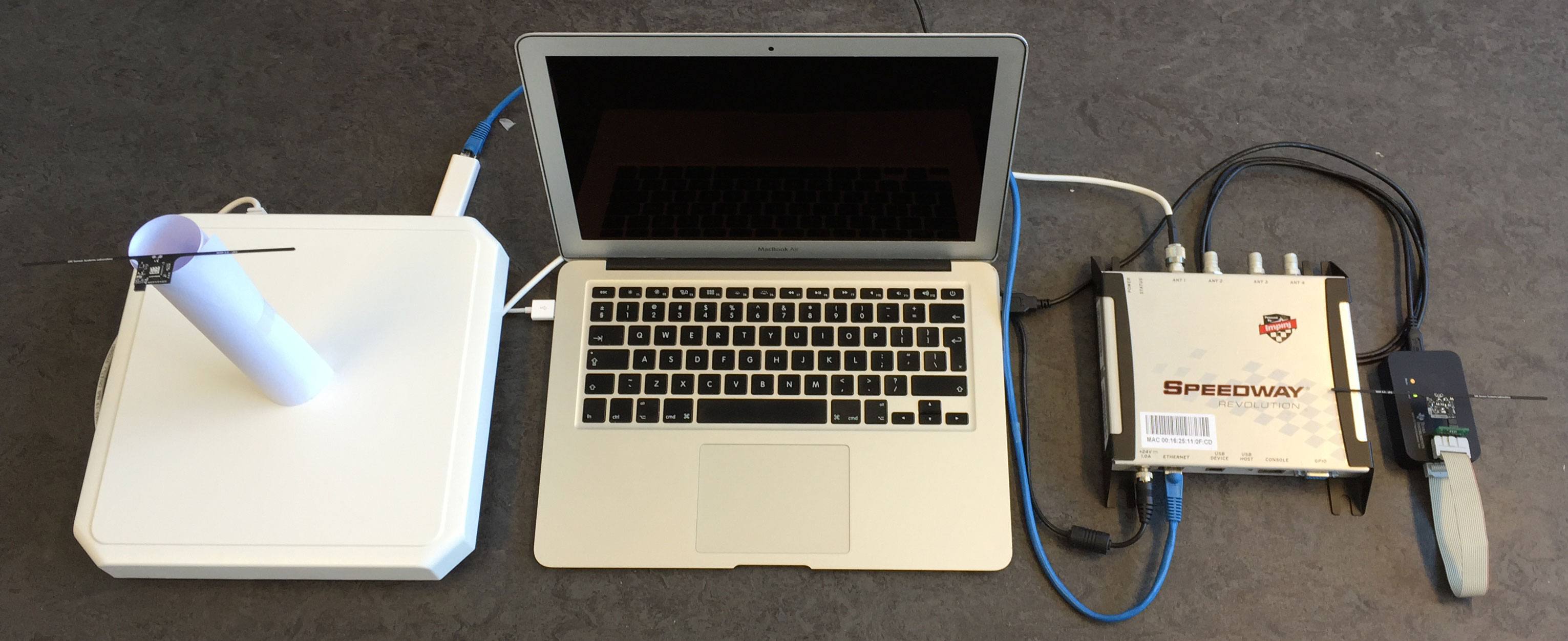}}
	\subfigure[Protocol layers]{
		\begin{tikzpicture}[-,>=stealth', semithick, node distance=15mm, scale=1.4, every node/.style={scale=1.4}]
		\scriptsize
			\draw[use as bounding box, transparent] (-1.4,-2.3) rectangle (1.6,0.3);
			
			\node [a] (A)                            {Host (PC)};
			\node [a] (B) at ([shift={(0,-1)}] A)    {Reader};
			\node [a] (C) at ([shift={(0,-1)}] B)  {Tag};
			
			\draw [draw=black,thin] ([shift={(0,0.12)}]B.east) -- ([shift={(0.3,0.12)}]B.east) |- ([shift={(0.3,0.7)}]B.east);
			\draw [draw=black,thin] ([shift={(0.3,0.4)}]B.east) -- ([shift={(0.4,0.7)}]B.east);
			\draw [draw=black,thin] ([shift={(0.3,0.4)}]B.east) -- ([shift={(0.2,0.7)}]B.east);
			
			\draw [draw=black,thin] ([shift={(0,0)}] C.west) -- ([shift={(-0.3,0)}]C.west) |- ([shift={(-0.3,0.7)}]C.west);
			\draw [draw=black,thin] ([shift={(-0.3,0.4)}]C.west) -- ([shift={(-0.4,0.7)}] C.west);
			\draw [draw=black,thin] ([shift={(-0.3,0.4)}]C.west) -- ([shift={(-0.2,0.7)}] C.west);
			
			\draw [draw=black] (A.south) -- (B.north);
			\draw [<->, draw=black,thin] (A.west) -- ([shift={(-0.3,0)}]A.west) |- ([shift={(-0.3,0)}]B.west) -- (B.west);
			\draw [<->, draw=black,thin] (B.east) -- ([shift={(0.3,0)}]B.east) |- ([shift={(0.3,0)}]C.east) -- (C.east);
			
			\node [thick, rotate=90] at (-1,-0.45) [above] {LLRP};
			\node [thick, rotate=-90] at (1,-1.5) [above] {EPC C1G2};
			
		\end{tikzpicture}
	}
\vspace{-2.5mm}
\caption{CRFID system setup: (a) hardware (from left to right) (i) paper tube-protruded WISP\,5 on an antenna, (ii) host (PC), (iii) reader---connected to an antenna and a host, (iv) Texas Instruments MSP430 Flash Emulation Tool connected to a WISP and a host; (b) Protocol layers of CRFID communication.}
\label{fig:systemsetup}
\vspace{-2.5mm}
\end{figure}
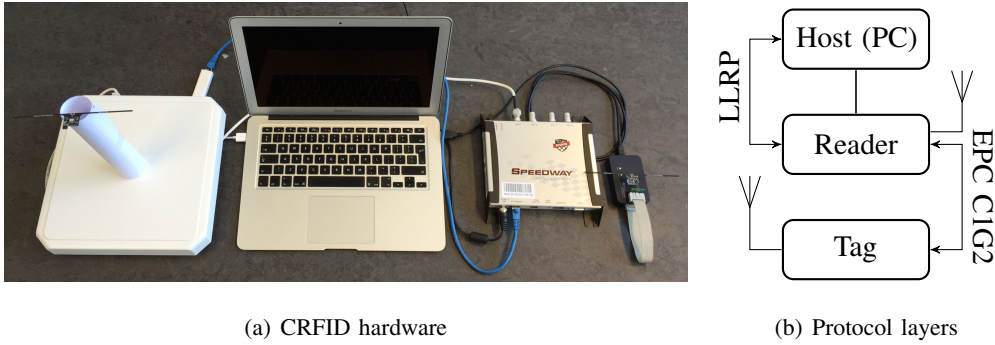

\subsection{Reader-to-Tag Communication: EPC C1G2}
\label{sec:epcgen2}

There are only two commands in the EPC C1G2 standard with the purpose of sending data from the reader to a tag: \verb|Write| and \verb|BlockWrite|~\cite{epc2015}.
While \verb|Write| is a mandatory command in the EPC C1G2 specification (i.e. tags conforming to the standard must support these commands~\cite[pp. 13]{epc2015}), \verb|BlockWrite| is specified as optional, and prior to this work its implementation was nonexistent in CRFIDs.
Data transfer with \verb|Write| is limited to only one word at a time.
To achieve data transmission rates beyond what is supported with \verb|Write|, an implementation of \verb|BlockWrite| for CRFIDs is fundamentally needed. Length-wise, \verb|BlockWrite| can be orders of magnitude longer than \verb|Write|. It is thus far more susceptible to channel errors (induced by e.g. CRFID movement). Therefore its performance must be well understood to maximize the transfer speeds for CRFID downstream protocols built on top of EPC C1G2.

\subsection{Host-to-Reader Communication: LLRP}
\label{sec:llrp}

Despite LLRP's large overhead, it is possible to use it as part of a downstream protocol.

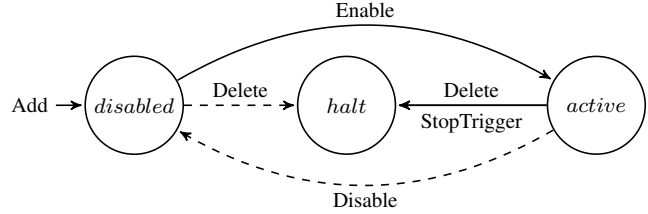
\begin{figure}
	\centering
	\scriptsize
	\begin{tikzpicture}[->,>=stealth',shorten >=1pt, semithick, scale=1.0, every node/.style={scale=1.0}]
		\node[initial,state,initial text={Add}, minimum size=13mm] (A)                          {\scriptsize{$disabled$}};
		\node[state]                         (C) [right = 15mm of A, minimum size=13mm]      {\scriptsize{$halt$}};
		\node[state]                         (B) [right = 20mm of C, minimum size=13mm]      {\scriptsize{$active$}};
		
		\path (A) edge [bend left] node[midway,above] {Enable} (B);
		\path [dashed] (A) edge             node[midway,above] {Delete} (C);
		\path (B) edge             node[midway,above] {Delete} (C);
		\path (B) edge             node[midway,below] {StopTrigger} (C);
		\path [dashed] (B) edge [bend left] node[midway,below] {Disable} (A);
	\end{tikzpicture}
	\vspace{-3mm}
	\caption{LLRP AccessSpec reader state machine~\cite[Fig. 10]{epc2010} utilized in enabling continuous host-to-reader bulk data transfer following Proposition 1. The dashed lines represent optional state transitions.}
	\label{fig:llrp}
	\vspace{-2.5mm}
\end{figure}

\begin{proposition}Continuous host-to-reader data streaming is enabled in LLRP by issuing a train of \verb|AccessSpec| commands.\begin{evidence} A host machine can command a reader to start an inventory session on tags only by enabling a \verb|ROSpec|~\cite[Sec. 6]{epc2010}.
An \verb|AccessSpec| can optionally be included inside a \verb|ROSpec| to make a reader issue an access command (e.g. \verb|Write|) on tags.
To change the current command or its properties (e.g. the amount of words or content of a \verb|BlockWrite|) issued by the reader, the \verb|AccessSpec| must be changed.
To achieve this, the host machine needs to tell the reader to first delete the current \verb|AccessSpec|, add a new one, and finally enable it~(see~\Cref{fig:llrp}).
Alternatively, an \verb|AccessSpecStopTrigger| can be specified by the host to make a reader autonomously delete an \verb|AccessSpec| once its command has been performed \verb|OperationCountValue| times by the tag.
Using both principles, it is possible for a host to send multiple \verb|AccessSpecs| in a continuous fashion to a reader to effectively enable a data stream.\hfill $\square$\end{evidence}
\end{proposition}

\subsection{Design Requirements}

From the challenges that arise in~\Cref{sec:introduction} and the limitations found in CRFID systems, we can now formulate the design requirements for a downstream CRFID protocol.
It must be \begin{inparaenum}[\itshape(i)]\item built on top of EPC C1G2 and LLRP; \item able to send large portions of data from a host to a CRFID via a reader; 
data transmission and storage must be reliable and \item tolerate interruptions to operating power; \item must tolerate changes of distance of the CRFID platform to the reader's antenna.
Additionally, to achieve transmission rates beyond the limitations set by EPC C1G2 \verb|Write|, \item the \verb|BlockWrite| command should be enabled.
Furthermore, to alleviate the negative effect of transient power on repetitive tasks in write operations to CRFID's memory, such protocol must use \item CRFID platforms with non-volatile FRAM memory\end{inparaenum}.
We remark that although a CRFID system might consist of multiple tags, we design Wisent for communication with a single tag in mind\footnote{Extensions of Wisent for communication with multiple tags will be discussed in~\Cref{sec:futurework}.}.


\section{Wisent: Design and Implementation}
\label{sec:implementation}

\subsection{Wisent Hardware and Software}
\label{sec:wisent_hardware}

As a host we use a x86-64 computer with an Intel i5-3317U processor running Ubuntu 14.04.
As an RFID reader we use an off-the-shelf 915\,MHz Impinj Speedway\footnote{We have also tested Wisent using Impinj Speedway R1000 but the experimental results in this work were generated with the R420 reader only.} R420 with firmware version 4.8.3 connected to a Laird S9028PCR 8.5\,dBic gain antenna.
As a CRFID device we use a WISP\,5~\cite{wisp5} with TI MSP430FR5969 MCU which has 64\,KB of FRAM\footnote{Although Wisent has been tested exclusively on the WISP\,5, it can be targeted towards any CRFID platform using FRAM non-volatile memory.}.
To initially program the WISP with Wisent we use a MSP430 Flash Emulation Tool (FET), in combination with TI Code Composer Studio (CCS), attached to the host.
The FET also enables measuring the energy consumption of the WISP and inspecting WISP non-volatile memory.

To implement all necessary features of Wisent in LLRP we have extended sllurp~\cite{sllurp}: a LLRP control library written in Python.
The CRFID-side of Wisent is implemented in C and MSP430 assembly. 
A complete Wisent implementation amasses to 350 lines of Python and 60 lines of C in addition to 100 instructions of MSP430 assembly.
Refer again to \Cref{fig:systemsetup} for our CRFID system setup used in Wisent experiments.

\subsection{Wisent Components}

\subsubsection{Message Format}

To implement fast reader-to-tag transmission, any bulk data that is to be sent to CRFID (e.g. regular file or firmware) should be presented in one specific format.
For that purpose we utilize the Intel Hex file format due to its popularity, noting that other formats such as TekHex or Motorola-S~\cite[Sec. 12.12]{slau131} can be also supported.
An Intel Hex file is divided into lines, called records, and contain fields of information about the data, see~\Cref{fig:ihex}, which are later parsed for transfer as described below.

\newcommand{\omitmidbitsihex}[1]{%
	\tiny
	\ifnum#1=1234567890
		#1
	\else
		\ifnum#1>14
			\ifnum#1=15
				$n$-$1$
			\else
				$n$
			\fi
		\else
			\ifnum#1<11
				#1
			\else
				\ifnum#1=12
					$\dots$
				\fi
			\fi
		\fi
	\fi
}
\newcommand{\colorbitbox}[3]{%
	\rlap{\bitbox{#2}{\color{#1}\rule{\width}{\height}}}%
\bitbox{#2}{#3}}

\begin{figure}
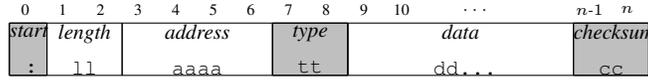

	\begin{center}
	\scriptsize
	\begin{bytefield}[boxformatting={\centering\itshape}, bitformatting=\omitmidbitsihex, bitwidth=5mm, endianness=little]{17}
	\bitheader{0-16}\\
	\colorbitbox{lightgray}{1}{start\\ \tt :}                   &
	\bitbox{2}{length\\ \tt ll}                 &
	\bitbox{4}{address\\ \tt aaaa}              &
	\colorbitbox{lightgray}{2}{type\\ \tt tt}                   &
	\bitbox{6}{data\\ \tt dd\dots}            &
	\colorbitbox{lightgray}{2}{checksum\\ \tt cc}
	\end{bytefield}
	\end{center}
	\vspace{-2.5mm}
	\caption{Example record of an Intel Hex file. The numbers on top of each field denote the index of the ASCII symbol in the record and the values inside each field denote example content. The length field specifies the byte count in the data field, while the address field represents the destination address for the data in the CRFID memory. Gray fields are not used in Wisent messages.}
	\label{fig:ihex}
	\vspace{-2.5mm}
\end{figure}

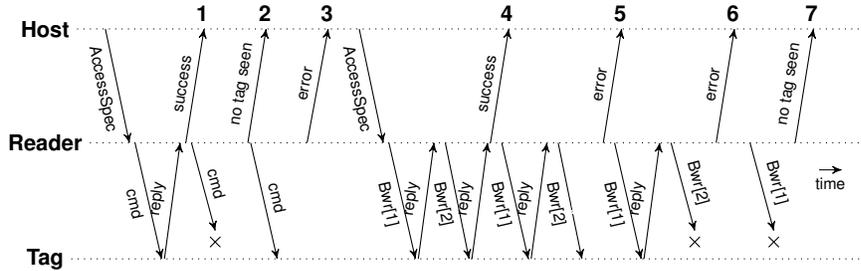
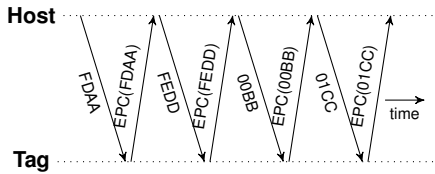
\begin{figure*}
\subfigure[EPC C1G2: standard versus Impinj readers; cmd: command (generic), Bwr{[}x{]}: $x$th\texttt{BlockWrite}]{
\centering
\begin{tikzpicture}[font=\sffamily,>=stealth',commentl/.style={text width=3cm},scale=0.75, every node/.style={scale=0.75}]
\draw[use as bounding box, transparent] (-0.7,-4.2) rectangle (14.5,0.4);
\node[] (init)                   {\textbf{Host}};
\node[below=11mm of init] (recv) {\textbf{Reader}};
\node[below=11mm of recv] (tag)  {\textbf{Tag}};

\draw[->] ([xshift=5mm]init.east)   coordinate (msg1u)     -- ([xshift=7mm]recv.east) coordinate (msg1m) node[midway, below, sloped]     {\scriptsize AccessSpec};
\draw[->] ([xshift=1mm]msg1m)       coordinate (msg2m)     -- ([xshift=16mm]tag.east) coordinate (msg2s) node[midway, below, sloped]     {\scriptsize cmd};
\draw[->] ([xshift=0.5mm]msg2s)     coordinate (msg3s)     -- ([xshift=8mm]msg2m) coordinate (msg3m) node[midway, above, sloped]        {\scriptsize reply};
\draw[->] ([xshift=1mm]msg3m)       coordinate (msg4m)     -- ([xshift=17.5mm]msg1u) coordinate (msg2u) node[midway, above, sloped]      {\scriptsize success};
\draw[->] ([xshift=2mm]msg3m)       coordinate (msg5m)     -- ([xshift=9mm, yshift=5mm]msg3s) coordinate (msg4s) node[midway, above, sloped]        {\scriptsize cmd};
\draw ([yshift=-2mm]msg4s) node[cross] {};
\draw[->] ([xshift=10mm]msg5m)      coordinate (msg6m)     -- ([xshift=11mm]msg2u) coordinate (msg3u) node[midway, above, sloped]      {\scriptsize no tag seen};
\draw[->] ([xshift=0.5mm]msg6m)      coordinate (msg7m)     -- ([xshift=20mm]msg3s) coordinate (msg5s) node[midway, above, sloped]      {\scriptsize cmd};
\draw[->] ([xshift=10mm]msg7m)      coordinate (msg8m)     -- ([xshift=11mm]msg3u) coordinate (msg4u) node[midway, above, sloped]      {\scriptsize error};

\node[commentl, above right = 0mm and -2mm of msg2u.north]{\textbf{1}};
\node[commentl, above right = 0mm and -2mm of msg3u.north]{\textbf{2}};
\node[commentl, above right = 0mm and -2mm of msg4u.north]{\textbf{3}};

\draw[->] ([xshift=45mm]msg1u)   coordinate (msg1u)     -- ([xshift=45mm]msg1m) coordinate (msg1m) node[midway, below, sloped]     {\scriptsize AccessSpec};
\draw[->] ([xshift=45mm]msg2m)   coordinate (msg2m)     -- ([xshift=45mm]msg2s) coordinate (msg2s) node[midway, below, sloped]     {\scriptsize Bwr[1]};
\draw[->] ([xshift=45mm]msg3s)   coordinate (msg3s)     -- ([xshift=45mm]msg3m) coordinate (msg3m) node[midway, above, sloped]        {\scriptsize reply};
\draw[->] ([xshift=45mm]msg5m)   coordinate (msg5m)     -- ([xshift=45mm,yshift=-5mm]msg4s) coordinate (msg4s) node[midway, below, sloped]        {\scriptsize Bwr[2]};
\draw[->] ([xshift=9.5mm]msg3s)  coordinate (msg5s)     -- ([xshift=9.5mm]msg3m) coordinate (msg4m) node[midway, above, sloped]        {\scriptsize reply};
\draw[->] ([xshift=43mm]msg6m)   coordinate (msg6m)     -- ([xshift=43mm]msg3u) coordinate (msg3u) node[midway, above, sloped]      {\scriptsize success};
\node[commentl, above right = 0mm and -2mm of msg3u.north]{\textbf{4}};

\draw[->] ([xshift=20mm]msg2m)   coordinate (msg2m)     -- ([xshift=20mm]msg2s) coordinate (msg2s) node[midway, below, sloped]     {\scriptsize Bwr[1]};
\draw[->] ([xshift=20mm]msg3s)   coordinate (msg3s)     -- ([xshift=20mm]msg3m) coordinate (msg3m) node[midway, above, sloped]        {\scriptsize reply};
\draw[->] ([xshift=20mm]msg5m)   coordinate (msg5m)     -- ([xshift=20mm]msg4s) coordinate (msg4s) node[midway, below, sloped]        {\scriptsize Bwr[2]};
\draw[->, white] ([xshift=20mm]msg5s)   coordinate (msg5s)     -- ([xshift=20mm]msg4m) coordinate (msg4m) node[midway, above, sloped, white]        {\scriptsize reply};
\draw[->] ([xshift=20mm]msg6m)   coordinate (msg6m)     -- ([xshift=20mm]msg3u) coordinate (msg3u) node[midway, above, sloped]      {\scriptsize error};
\node[commentl, above right = 0mm and -2mm of msg3u.north]{\textbf{5}};

\draw[->] ([xshift=20mm]msg2m)   coordinate (msg2m)     -- ([xshift=20mm]msg2s) coordinate (msg2s) node[midway, below, sloped]     {\scriptsize Bwr[1]};
\draw[->] ([xshift=20mm]msg3s)   coordinate (msg3s)     -- ([xshift=20mm]msg3m) coordinate (msg3m) node[midway, above, sloped]        {\scriptsize reply};
\draw[->] ([xshift=20mm]msg5m)   coordinate (msg5m)     -- ([xshift=20mm,yshift=5mm]msg4s) coordinate (msg4s) node[midway, above, sloped]        {\scriptsize Bwr[2]};
\draw ([yshift=-2mm]msg4s) node[cross] {};
\draw[->, white] ([xshift=29mm]msg5s)   coordinate (msg5s)     -- ([xshift=29mm]msg4m) coordinate (msg4m) node[midway, above, sloped, white]        {\scriptsize reply};
\draw[->] ([xshift=20mm]msg6m)   coordinate (msg6m)     -- ([xshift=20mm]msg3u) coordinate (msg3u) node[midway, above, sloped]      {\scriptsize error};
\node[commentl, above right = 0mm and -2mm of msg3u.north]{\textbf{6}};

\draw[->] ([xshift=14mm]msg5m)   coordinate (msg5m)     -- ([xshift=14mm]msg4s) coordinate (msg4s) node[midway, above, sloped]        {\scriptsize Bwr[1]};
\draw ([yshift=-2mm]msg4s) node[cross] {};
\draw[->] ([xshift=14mm]msg6m)   coordinate (msg6m)     -- ([xshift=14mm]msg3u) coordinate (msg3u) node[midway, above, sloped]      {\scriptsize no tag seen};
\node[commentl, above right = 0mm and -2mm of msg3u.north]{\textbf{7}};

\draw[dotted, shorten >=-1cm] (init) -- ([xshift=128mm]init.east);
\draw[dotted, shorten >=-1cm] (recv) -- ([xshift=125mm]recv.east);
\draw[dotted, shorten >=-1cm] (tag)  -- ([xshift=128mm]tag.east);
\draw[->] ([yshift=-25mm, xshift=1mm]msg3u) coordinate (leftright2l) -- ([yshift=-25mm, xshift=5mm]msg3u) coordinate (leftright2r) node[midway, below] {\scriptsize time};

\end{tikzpicture}\label{fig:readerinternals}}
\subfigure[Wisent Basic: \!\! \texttt{:02AADD00BBCCF0} record]{\centering
\begin{tikzpicture}[font=\sffamily,>=stealth', commentl/.style={text width=3cm},scale=0.75, every node/.style={scale=0.75}]

\node[] (init)                  {\textbf{Host}};
\node[below=1.5cm of init] (recv) {\textbf{Tag}};

\draw[->] ([xshift=0.3cm]init.east)     coordinate (msg1u)     -- ([xshift=1.2cm]recv.east) coordinate (msg1d) node[midway, below, sloped] {\scriptsize FDAA};
\draw[->] ([xshift=0.1cm]msg1d)         coordinate (msg2d)     -- ([xshift=1.3cm]msg1u) coordinate (msg2u) node[midway, above, sloped]     {\scriptsize EPC(FDAA)};
\draw[->] ([xshift=0.1cm]msg2u)         coordinate (msg3u)     -- ([xshift=1.3cm]msg2d) coordinate (msg3d) node[midway, below, sloped]     {\scriptsize FEDD};
\draw[->] ([xshift=0.1cm]msg3d)         coordinate (msg4d)     -- ([xshift=1.3cm]msg3u) coordinate (msg4u) node[midway, above, sloped]     {\scriptsize EPC(FEDD)};
\draw[->] ([xshift=0.1cm]msg4u)         coordinate (msg5u)     -- ([xshift=1.3cm]msg4d) coordinate (msg5d) node[midway, below, sloped]     {\scriptsize 00BB};
\draw[->] ([xshift=0.1cm]msg5d)         coordinate (msg6d)     -- ([xshift=1.3cm]msg5u) coordinate (msg6u) node[midway, above, sloped]     {\scriptsize EPC(00BB)};
\draw[->] ([xshift=0.1cm]msg6u)         coordinate (msg7u)     -- ([xshift=1.3cm]msg6d) coordinate (msg7d) node[midway, below, sloped]     {\scriptsize 01CC};
\draw[->] ([xshift=0.1cm]msg7d)         coordinate (msg8d)     -- ([xshift=1.3cm]msg7u) coordinate (msg8u) node[midway, above, sloped]     {\scriptsize EPC(01CC)};

\draw[dotted, shorten >=-1cm] (init) -- (msg7d|-init);
\draw[dotted, shorten >=-1cm] (recv) -- (msg7d|-recv);
\draw[->] ([yshift=-1.5cm, xshift=12mm]msg7u) coordinate (leftright2l) -- ([yshift=-1.5cm, xshift=6mm]msg8u) coordinate (leftright2r) node[midway, below] {\scriptsize time};
\end{tikzpicture}
\label{fig:wisentwrite}}
\cprotect\caption{Wisent message transfer case studies: (a) Difference of using EPC C1G2 commands as specified by the standard~\cite{epc2015} (\textbf{1}--\textbf{3}) and \texttt{BlockWrite} with \verb|WordCount=2| as observed at Impinj readers (\textbf{4}--\textbf{7}).
The reader reports a no tag seen (\textbf{2},\textbf{7}) if the tag did not receive the command at all.
An error is reported instead if the tag received the command but processed it incorrectly (\textbf{3},\textbf{5},\textbf{6}).
Only success reports (\textbf{1},\textbf{4}) are counted towards the \verb|OperationCountValue| of individual \verb|AccessSpecs|; (b) Example of a Wisent Basic communication sequence of messages constructed from a record with content \verb+:02AADD00BBCCF0+.}
\end{figure*}

\subsubsection{Message Definitions}
\label{sssec:messagedef}

We define $m$: a \emph{Wisent message}, i.e. message sent from the host to CRFID as content for a \verb|Write| or \verb|BlockWrite| command that is constructed by the host as follows.
First, information is taken from the length, address and data fields of each Intel Hex file record $i$, being the only necessary fields for our implementation.
We denote the data that is taken this way as a matrix $I$, from which $I(i,j)$ is the $j$-th chunk in row $i$ of size $S_{\text{p}}$ words.
The value of $S_{\text{p}}$ depends on which of the two commands, leveraged by the two distinct versions of Wisent described in subsequent~\Cref{sec:wrprotocol} and~\Cref{sec:bwrprotocol}, the content is constructed for.
To a constructed $I(i,j)$, we add a header with information to identify and instruct the CRFID as to how the remaining part of the message, i.e. the payload, should be handled.
Details on the header and payload content, including payload handling, shall also be discussed in~\Cref{sec:wrprotocol} and~\Cref{sec:bwrprotocol}, as they are also Wisent-version dependent.

The next message to be sent is defined as
\begin{equation}
\label{eq:nextmbasic}
m_{\text{next}} =
	\begin{cases}
		I(1,1)     & \quad \text{if } m = \varepsilon,      \\
		I(i,j+1)   & \quad \text{if } j \neq \text{EOL}, \\
		I(i+1,1)   & \quad \text{if } j = \text{EOL},    \\
		\emptyset  & \quad \text{if } i = \text{EOF},                  \\
	\end{cases}
\end{equation}
where $\varepsilon$ is the undefined message and EOL and EOF is the Intel Hex end of the record and end of file, respectively.

The backscattered \verb|EPC| field of the CRFID is utilized for message verification purposes.
During the construction of each message, the host generates verification data that is compared with the \verb|EPC| in a tag report from the reader.
However, since the \verb|EPC| is backscattered before a command that accesses the tag is executed~\cite[Annex E]{epc2015} (e.g. \verb|Write|, \verb|BlockWrite|),
an \verb|EPC| after that command is needed to verify the current message.
The host receives a \emph{positive acknowledgment} (ACK) of a message if this \verb|EPC| matches the verification data
and a \emph{negative acknowledgement} (NACK) otherwise. 

The \verb|OperationCountValue| (\verb|OCV|), described in~\Cref{sec:llrp}, acts as an upper bound of the \emph{operation frame}~\cite[Fig. 4.3]{tan2015:thesis} in which a message $m$ is actively transmitted by the reader through a command operation.
Unlike ACKs and NACKs, which depend on following \verb|EPC| fields, the result of a command operation is reported before the next \verb|EPC| arrives, and only successful commands count towards the \verb|OCV|, see~\Cref{fig:readerinternals}.
However, such a report contains an \verb|EPC| regardless of result and therefore, the report of a successful operation, e.g. \verb|Write|, can contain a NACK and contrarily a report of an erroneous operation can embed an ACK\footnote{EPC of previously reported (successful) operation is piggybacked \emph{during the next operation} (which happens to be unsuccessful) as an ACK.}.

The \emph{message frame} of a message $m$ starts as soon as the host sends that message to the reader.
When $m$ is acknowledged by the CRFID, the message frame of $m$ ends and the host moves on to the message frame of $m_{\text{next}}$,
commanding the reader to delete the \verb|AccessSpec| containing $m$ before sending $m_{\text{next}}$.
However, there is a delay before this deletion is processed by the reader and thus a delay before the start of the operation frame of $m_{\text{next}}$.
Because of this, the operation frame of $m$ overlaps with the message frame of $m_{\text{next}}$ causing ACKs for $m$ in the overlap to be NACKs for $m_{\text{next}}$.

If too many NACKs, defined by $N_{\text{threshold}}$, are observed, a \emph{timeout} is generated and the message will be resent up to a maximum, $R_{\text{max}}$, number of times.
If after the maximum amount of resends, the message is still not received, the Wisent transfer is aborted and ends up in a failure.
Even though power failures as a result of transient power (see~\Cref{fig:powercycles}) are accounted for by NACKs, separation of the CRFID from the antenna for a prolonged period of time (which results in a transfer failure) are not.
This issue will be discussed in~\Cref{sec:futurework}.

\subsection{Wisent: Basic Protocol}
\label{sec:wrprotocol}


\floatname{algorithm}{Protocol}
\renewcommand{\algorithmicfor}{\textbf{upon}}
\renewcommand{\algorithmicdo}{}

\setlength{\textfloatsep}{2pt}%
\begin{algorithm}[t]
	\caption{Wisent Basic}
	\scriptsize
	\label{protocol1}
	\begin{algorithmic}[1]
		\State $R_{\text{count}} \leftarrow 0$
		\Comment{\textbf{Host events}}
		\State $m \leftarrow m_{\text{next}}$
		\Comment{See~\cref{eq:nextmbasic}}
		\State \textsc{send}($m$)
		\While{$R_{\text{count}} < R_{\max}$}
		\vspace{1mm}
		\For{\textsc{ack}:}
		\State $R_{\text{count}} \leftarrow 0$
		\State $m \leftarrow m_{\text{next}}$
		\Comment{See~\cref{eq:nextmbasic}}
		\State \textsc{send}($m$)
		\EndFor
		\For{\textsc{timeout}:}
		\State $R_{\text{count}} \leftarrow R_{\text{count}}+1$
		\State \textsc{send}($m$)
		\EndFor
		\EndWhile
		\For{\textsc{receive}($m$):}
		\Comment{\textbf{Tag events}}
			\State $\texttt{EPC} \leftarrow  \textsc{handle}(m)$\label{prot1:handle}
			\State \textsc{backscatter}(\texttt{EPC})
		\EndFor
	\end{algorithmic}
\end{algorithm}

The \verb|Write| command limits a message to only one word of content.
Due to the small size of the message, we propose a copy of the message to be kept by the host as verification data to compare with the \verb|EPC|.
By fitting a single byte header in such a message, only one byte remains for the payload, i.e. $S_{\text{p}} = 0.5$ words.
We propose a non-ambiguous header that describes the payload with data $I(i,j)$ obtained from only the address and data fields.
For each row $i$, two messages should be sent, each with a payload containing half of the address field and a unique header to identify which half of the address field the message contains.
The following messages should include a byte from the data field as payload and a header that represents an offset of the byte to the base address.
The proposed header definitions are: \begin{inparaenum}[\itshape(i)]\item \verb|FD|: new line (address low byte), \item \verb|FE|: address high byte, and \item \verb|00|--\verb|20| data byte with offset. \end{inparaenum} 
For ease of explanation, an example message transfer sequence is shown in~\Cref{fig:wisentwrite}.

Messages received by the CRFID are handled by a $\textsc{handle}(m)$ function, see Protocol~\ref{protocol1} (line \ref{prot1:handle}), as follows.
First, the integrity of a message is checked and the CRC16 checksum over data of the \verb|Write| command is calculated~\cite[Sec. 6]{epc2015}.
However, data written to non-volatile memory might still be corrupted in case of power failure.
Therefore, the address to which the byte was written is immediately read to verify the content.
Afterwards, the \verb|EPC| of the CRFID is set to the header of the received message along with the read byte.
Pseudocode of Wisent Basic can be found in Protocol~\ref{protocol1}.

\subsection{Wisent EX: Extending Wisent Basic with BlockWrite}
\label{sec:bwrprotocol}

\setlength{\textfloatsep}{2pt}%
\begin{algorithm}[t]
	\scriptsize
	\caption{Wisent EX}
	\label{protocol2}
	\begin{algorithmic}[1]
		\State $R_{\text{count}} \leftarrow 0$
		\Comment{\textbf{Host events}}
		\State $M_{\text{count}} \leftarrow 0$
		\State $S_{\text{p}} \leftarrow S_{\text{max}}$
		\State $m \leftarrow m_{\text{next}}$
		\Comment{See~\cref{eq:nextmbasic}}
		\State \textsc{send}($m$)
		\While{$R_{\text{count}} < R_{\text{max}}$}
			\For{\textsc{ack}:}
				\State $R_{\text{count}} \leftarrow 0$
				\If{$M_{\text{count}} > M_{\text{threshold}}$}
					\State $\textsc{throttle}(S_{\text{p}},\scriptsize\text{up})$
					\Comment{See~\cref{eq:throttleequation}}
					\State $M_{\text{count}} \leftarrow 0$
				\Else
					\State $M_{\text{count}} \leftarrow M_{\text{count}}+1$
				\EndIf
				\State $m \leftarrow m_{\text{next}}$
				\Comment{See~\cref{eq:nextmbasic}}
				\State \textsc{send}($m$)
			\EndFor
			\For{\textsc{timeout}:}
				\State $R_{\text{count}} \leftarrow R_{\text{count}}+1$
				\State $M_{\text{count}} \leftarrow 0$
				\State $\textsc{throttle}(S_{\text{p}},\scriptsize\text{down})$
				\Comment{See~\cref{eq:throttleequation}}
				\State $m \leftarrow I(i,j)$
				\Comment{See~\cref{eq:nextmbasic}}
				\State \textsc{send}($m$)
			\EndFor
		\EndWhile
		\For{\textsc{receive}($m$):}
		\Comment{\textbf{Tag events}}
			\State $\texttt{EPC} \leftarrow  \textsc{handle}(m)$
			\State \textsc{backscatter}(\texttt{EPC})
		\EndFor
	\end{algorithmic}
\end{algorithm}

To overcome the single word limit imposed by the \verb|Write| command, we have extended Wisent to make use of \verb|BlockWrite|, denoted as Wisent EX, which introduces the use of a \verb|WordCount| parameter.
\verb|WordCount| specifies the size of a \verb|BlockWrite| payload.

\subsubsection{Implementation of BlockWrite}
\label{sssec:bwrimplementation}

We have observed that the Impinj RFID readers do not issue the \verb|BlockWrite| command as specified by EPC C1G2~\cite[pp. 92]{epc2015}.
Rather than issuing one \verb|BlockWrite| command with data containing all the words,
the reader issues a \emph{series} of commands, see~\Cref{fig:readerinternals}, each containing one word and a sequential increasing address pointer to store the word in the memory of the tag\footnote{Although Wisent EX is built utilizing this non-standard \verb|BlockWrite| behavior, it would also work in combination with RFID readers that \emph{do} issue \verb|BlockWrite| as specified in the EPC C1G2 standard.}.
Because of this, the original \verb|BlockWrite| from the host to the reader is not known to the CRFID, while the reader still considers the series as the instructed \verb|BlockWrite|.
A command operation of \verb|BlockWrite| will only be reported as successful if and only if the CRFID processes each of the individual commands in the series and replied to the reader for each of these commands.
To avoid extra computation on the CRFID, the CRC16 checksum of each command in the series is not calculated.

\newcommand{\omitmidbitswisent}[1]{%
	\tiny
	\ifnum#1=1234567890
		#1
	\else
		\ifnum#1>5
			\ifnum#1=6
				$n$-$1$
			\else
				$n$
			\fi
		\else
			\ifnum#1<5
				#1
			\else
				\ifnum#1=5
					$\dots$
				\fi
			\fi
		\fi
	\fi
}

\begin{figure}
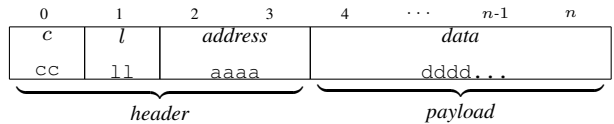

	\centering
	\scriptsize
	\vspace{2.5mm}
	\begin{bytefield}[boxformatting={\centering\itshape}, bitformatting=\omitmidbitswisent, bitwidth=10mm, endianness=little]{8}
	\bitheader{0-7}\\
	\bitbox{1}{$c$\\ \tt cc}                      &
	\bitbox{1}{$l$\\  \tt ll}               &
	\bitbox{2}{address\\ \tt aaaa}                &
	\bitbox{4}{data\\ \tt dddd\dots}            \\
	\bitbox[t]{0}{$\hspace{1mm}\underbrace{\hspace{38mm}}_{\text{\scriptsize header}}$}           &
	\bitbox[t]{0}{$\hspace{41mm}\underbrace{\hspace{38mm}}_{\text{\scriptsize payload}}$}
	\end{bytefield}
	\vspace{-2.5mm}
	\caption{Proposed Wisent EX message format. The numbers on top of each field denote the byte index.}
	\label{fig:wisentmessage}
\end{figure}

\subsubsection{Verification}

Because the CRC16 checksum was not calculated for \verb|BlockWrite|, the communication channel is not reliable anymore.
An alternative verification mechanism and checksum is necessary to secure the robustness of the communication channel and verify the data written to the non-volatile memory.
We propose the use of a single byte checksum for each message that is calculated by taking the least significant byte of the sum\footnote{Our approach follows the same method used in Intel Hex files but we are naturally aware of alternative error correction methods. Nevertheless, we will introduce another bootloader specific layer in~\Cref{sec:evalwireless}.} of all bytes in message $m$.
This checksum is calculated to check the integrity of the message when received by the CRFID and after writing data to the non-volatile memory.

With messages sizes that can exceed the \verb|EPC| length, using a copy of the message as means for verification is no longer a viable solution.
Nevertheless, a larger message size allows a header of more than one byte, that can be used instead for verification purposes and should be set as \verb|EPC| by the CRFID.
In such a header, we propose to include (i) the message checksum $c$; (ii) length of the payload in bytes  $l$; and (iii) destination address of the payload.
Consequently, the payload should contain data $I(i,j)$ obtained only from the data field of Intel Hex file record, see~\Cref{fig:wisentmessage}.

\subsubsection{Throttling}
\label{sssec:throttling}

We first introduce the following observation.

\begin{lemma}
Larger \verb|WordCount| does not always correspond to faster bulk transfer rates.\begin{IEEEproof}
We prove this observation by contradiction. 
For a \verb|BlockWrite| command of length $L$\,bits the transmission overhead is~\cite[Table 6.43]{epc2015} $H=51$\,bits. 
Assuming a general complementary error function relating bit error to distance $d$~\cite[Eq. (3)]{lettieri:infocom:1998} $p_e(d)=\mathrm{erfc}(1/d)$ a normalized \verb|BlockWrite| throughput is $T(L,d)={L}/(L+H)(1-p_e(d)^{L+H}$ (compare with~\cite[Eq. (1)]{lettieri:infocom:1998}). 
Now, for \verb|BlockWrite| commands with data lengths $L_1=16$\,bytes and $L_2=32$\,bytes we have $T(L_1,d)<T(L_2,d)$ for $d=0.2$, while for $d=0.5$ the opposite holds, which completes the proof.
\end{IEEEproof}\end{lemma}

Following the above observation we propose the use of a throttling mechanism to adjust $S_{\text{p}}$, the payload size of messages, which is initialized to a user defined value $S_{\text{max}}$.
Given the amount of words $S_{\text{r}}$ in data row $I(i)$, we construct a set $T_t$ for $S_{\text{p}}$ values as
\begin{equation}
	T_t = \left\{S_{\text{p}} : S_{\text{p}} = \left\lceil\frac{S_{\text{r}}}{n}\right\rceil\right\}\\,\quad n, S_{\text{r}}\in\mathbb{N}^{+}, n \leq S_{\text{r}},
\end{equation}
where $n$ represents the number of data chunks, and in turn messages, in which that row should be split.
The throttling function is then defined as
\begin{equation}
\label{eq:throttleequation}
	\textsc{throttle}(S_{\text{p}},\rho) = S_{\text{p+1}} \in T_t,
\end{equation}
where $\rho=\{\text{up},\text{down}\}$, with $S_{\text{p+1}} > S_{\text{p}}$ if throttling up, and $S_{\text{p+1}} < S_{\text{p}}$ if throttling down.

If the need for a resend is perceived, $S_{\text{p}}$ is throttled down to decrease load on the communication channel.
On the other hand, $S_{\text{p}}$ is increased upon $M_{\text{threshold}}$, a threshold of consecutive successful messages.
With this, the full description of the protocol is complete, refer to pseudocode in Protocol~\ref{protocol2}.


\section{Experiment Results}
\label{sec:evaluation}

\begin{figure*}
	\vspace{-3mm}
	\centering
	\subfigure[SOPS ($\psi_{\text{s}}$)]{\includegraphics[width=0.65\columnwidth]{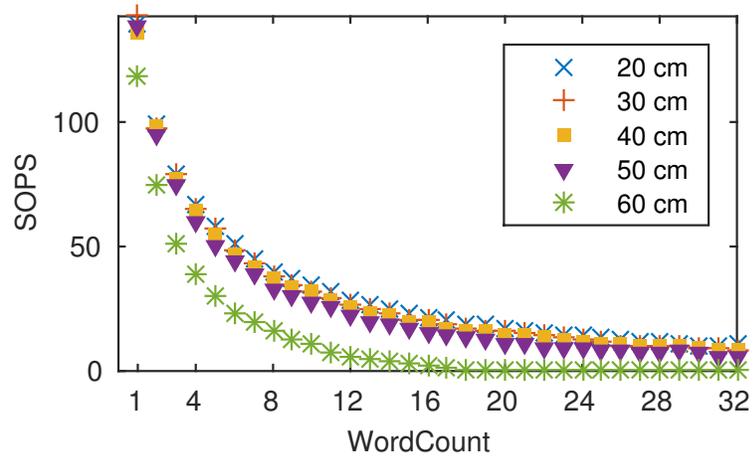}\label{fig:bwreval_ops}}
	\subfigure[Efficiency ($\eta$)]{\includegraphics[width=0.65\columnwidth]{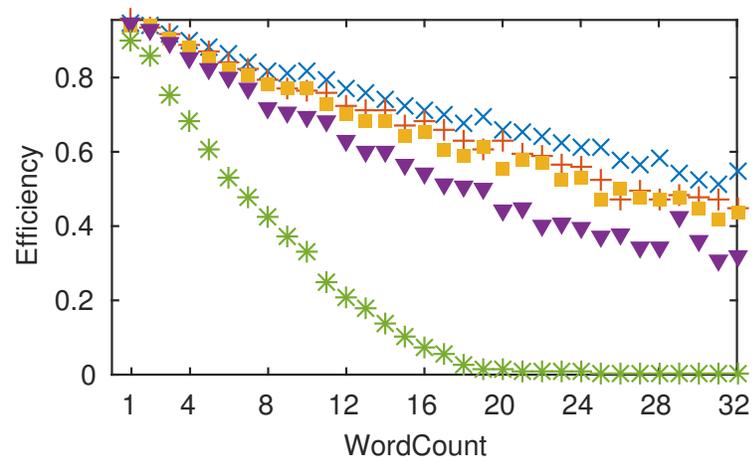}\label{fig:bwreval_efficiency}}
	\subfigure[Throughput ($\theta$)]{\includegraphics[width=0.65\columnwidth]{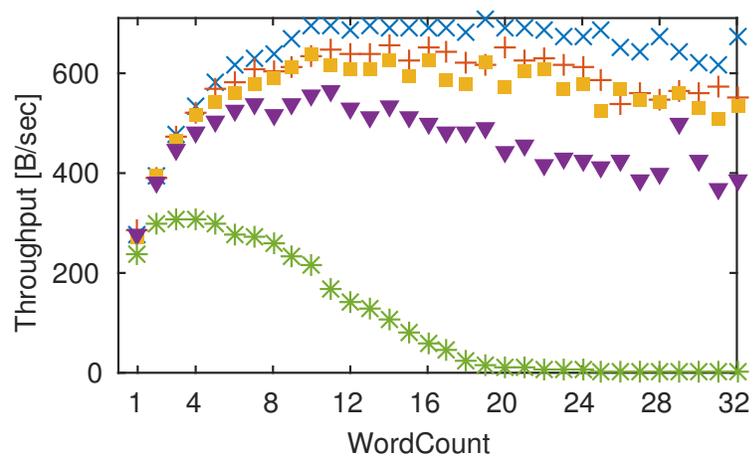}\label{fig:bwreval_throughput}}
	\vspace{-2.5mm}
	\caption{Success operations per second, efficiency and throughput of a single \texttt{BlockWrite} as functions of \texttt{WordCount} over various distances. }
	\label{fig:bwreval}
	\vspace{-4mm}
\end{figure*}

\subsection{Evaluation of BlockWrite}
\label{sec:evaluationofbwr}

We first evaluate the performance of the \verb|BlockWrite| alone, as it is the basis of Wisent. The experiment setup is as shown in~\Cref{fig:systemsetup} and explained in~\Cref{sec:wisent_hardware}. The whole measurement setup was located inside of an university laboratory and each of the experiments were repeated five times.

\subsubsection{Maximum Word Size of BlockWrite}
\label{sec:max_word_size}

EPC C1G2 implies a maximum value of 255\,words for \verb|WordCount| by the 8-bit length of the \verb|WordCount| header field. 
We have tested multiple \verb|WordCount| values and found a R420 reader is not able to issue more than 32 words at a time.

\subsubsection{BlockWrite Performance Metrics}
\label{sec:max_word_size}

During our experiments we let the reader issue a single \verb|BlockWrite| with a set \verb|WordCount| for each experiment. 
Such operation is repeated for a duration of ten seconds, i.e. the host sends an \verb|AccessSpec| with a \verb|BlockWrite| to the reader without an \verb|AccessSpecStopTrigger| and halts after ten seconds, over multiple distances $d=\{20,30,\ldots,60\}$\,cm from the tag to the antenna.

We propose the following set of performance criteria of the \verb|BlockWrite|: (i) number of \emph{success operations per second} (SOPS) defined as $\psi_{\text{s}} = {n_{\text{s}}}/{t}$ where $n_{\text{s}}$ is the number of success reports and $t$ (sec) is the duration of the experiment; (ii) the \emph{number of total operations per second} (TOPS) defined as $\psi_{\text{t}} = {n_{\text{t}}}/{t}$, (iii) \emph{efficiency} defined as $\eta = {\psi_{\text{s}}}/{\psi_{\text{t}}}$, and (iv) \emph{throughput} defined as $\theta = 2x\psi_{\text{s}}$\,(B/sec), where $x$ is the \verb|WordCount| value (since each word is 2 bytes long).

\subsubsection{BlockWrite Performance Results}
\label{sec:block_write_results}

In~\Cref{fig:bwreval_ops} we show that we are able to improve throughput of reader-to-CRFID communication manyfold compared to a single word length.
While there is only minor efficiency degradation between 20 and 50\,cm (\Cref{fig:bwreval_efficiency}), the effect of issuing \verb|BlockWrite| with large \verb|WordCount| can clearly be seen if the WISP is moved too far away from the antenna. The efficiency is much impacted larger at 60\,cm, possibly preventing further usage of such \verb|WordCount| values in Wisent EX messages.

\subsubsection{BlockWrite Performance Metrics Model}
\label{sec:block_write_results}

\begin{table}
	\centering
	\caption{Function parameters for $f_{\psi_{\text{t}}}(x)$ and $g_{\eta}(x)$ and $R^2$ for $h_{\eta}(x)$}
	\label{tab:funcparams}
	\vspace{-2.5mm}
	\begin{tabular}{| r || c | c || c | c | c || c |}
		\hline
		$d$ (cm) & $a_{\eta}$    & $b_{\eta}$ & $a_{2}$  & $b_{2}$ & $c_{2}$   & $R^{2}$ \\ \hline\hline
		20       & 0.0138        & 0.9448     & 170.3735 & 0.4184  & --22.1623  & 0.9176     \\ \hline
		30       & 0.0163        & 0.9401     & 166.3176 & 0.4523  & --16.6735  & 0.8627     \\ \hline
		40       & 0.0168        & 0.9270     & 164.6218 & 0.4341  & --18.7205  & 0.7716     \\ \hline
		50       & 0.0204        & 0.9056     & 158.3378 & 0.4909  & --10.9255  & 0.4504     \\ \hline
		60       & 0.0503        & 0.8710     & 122.2697 & 0.7347  &   11.0553  & 0.8553     \\ \hline
	\end{tabular}
\end{table}

From the measured data presented in~\Cref{fig:bwreval} we conjecture that simple functions can be used to describe all experimentally obtained statistics, which can be used in future analytical studies.
For this purpose we propose a model for which we introduce $f_{\psi_{\text{t}}}(x)$ and $g_{\eta}(x)$, which describe the relations between selected word size $x$ and $\psi_{\text{t}}$ and $\eta$, respectively. From these functions, $f_{\psi_{\text{s}}}(x)$ and $h_{\theta}(x)$ follow, describing the respective relations for $\psi_{\text{s}}$ and $\theta$.
We then have
\begin{equation}
f_{\psi_{\text{t}}}(x) = \frac{a_{2}}{x^{b_{2}}}+c_{2},
\end{equation}
\begin{equation}
g_{\eta}(x) = \frac{f_{\psi_{\text{s}}}(x)}{f_{\psi_{\text{t}}}(x)} = -a_{\eta}x + b_{\eta},
\end{equation}
\begin{equation}
 h_{\theta}(x) = 2xf_{\psi_{\text{s}}}(x),
\end{equation}
with all associated parameters given in~\Cref{tab:funcparams}. Using MATLAB R2015a we have calculated the above fit accuracy via coefficient of determination, $R^{2}$, for all fitted functions over all measured distances $d$.
For $f_{\psi_{\text{t}}}(x)$ the mean value of $R^{2}$ for all distances is $\mu_{R^{2}}(f_{\psi_{\text{t}}}(x))=0.9981$ with its variance of $\sigma_{R^{2}}^{2}(f_{\psi_{\text{t}}}(x))=5.423\times10^{-6}$ and $\mu_{R^{2}}(g_{\eta}(x))=0.9749$ with $\sigma_{R^{2}}^{2}(g_{\eta}(x))=1.8622\times10^{-4}$ indicating very good fit for the data.
The $R^{2}$ for $h_{\theta}(x)$ obtains lower accuracy due to outliers found in the measured data for $\theta$. Due to lower fit accuracy for $h_{\theta}(x)$ individual values of $R^2$ are given in~\Cref{tab:funcparams} for inspection.

\subsection{Evaluation of Wisent}
\label{sec:evaluation_wisent}

We now present the results for the complete Wisent protocol, but we shall only proceed with evaluation of Wisent EX due to its generic nature.
We will use the same experiment setup as in~\Cref{sec:evaluationofbwr}, unless stated otherwise.

As in the case of evaluating \verb|BlockWrite| performance we use tag-to-reader antenna distance $d$ as a parameter to evaluate Wisent.
However, instead of \verb|WordCount|, we take message payload size $S_{\text{p}}$ as the second parameter.

\subsubsection{Wisent Performance Metrics}
\label{sec:max_word_size}

In~\Cref{sec:evaluationofbwr} the experiments were executed with a single uninterrupted \verb|BlockWrite| of a set duration, which in Wisent is equivalent to a single message.
In Wisent, however, each of the previously used metrics also depends on the number of messages per second the RFID reader is able to process, i.e. the overhead discussed in~\Cref{sec:llrp}.
Furthermore, a Wisent log file provides information on events that occurred between messages, rather than events which occurred within a predefined time.
Therefore, we add $t$, the runtime of the transfer session, as a variable.

We introduce the following metrics: (i) the number of Wisent messages per second defined as $v = {m_{\text{t}}}/{t}$, where $m_{\text{t}}$ is the total amount of messages sent during the Wisent transfer session; (ii) number of success operations per message (SOPM) defined as $\psi_{\text{sm}}= {n_{\text{s}}}/{m_{\text{t}}}$, where $\psi_{\text{s}} = v \psi_{\text{sm}}$; (iii) the number of total operations per message defined as $\psi_{\text{tm}} = {n_{\text{t}}}/{m_{\text{t}}}$ with $\psi_{\text{t}} = v \psi_{\text{sm}}$; (iv) throughput defined as $\theta = 2S_{\text{p}} v$\,(B/sec), (v) the resend rate defined as $p_{\text{r}} = \frac{m_{\text{r}}}{m_{\text{r}} + m_{\text{s}}} = \frac{m_{\text{r}}}{m_{\text{t}}}$ where $m_{\text{r}}$ and $m_{\text{s}}$ are the number of messages resent and sent, respectively, and (vi) $\eta$, whose definition is given in~\Cref{sec:evaluationofbwr}.

\subsubsection{Wisent Performance Results}
\label{sec:wisent_performance_results}

We have experimented sending as many messages as possible and observed a value for $v\approx3.8$ messages per second when using the Impinj R420 RFID reader and only half that amount with an Impinj R1000.
This result is not affected by setting different values for \verb|OCV|, which should change the size of the operation frame, and in turn lead to a smaller or bigger message frame.
This has been confirmed by testing multiple values for $\texttt{OCV} = \{5,10,15,20,25,30\}$.
For $\texttt{OCV} = \{5,10\}$, the host is observed occasionally not receiving an ACK in each message frame before the operation frame of that message ends and the message is deleted.
For $\texttt{OCV} = 20$, the deletion of a message commanded by the host, explained in~\Cref{sssec:messagedef}, collides with the \verb|AccessSpecStopTrigger| after executing 20 operations causing the reader to misbehave and cease the bitstream.
Higher values for \verb|OCV| causes the next message frame of each message to be flooded with NACKs and forces the resend of the message.
Only for the value $\texttt{OCV} = 15$, the bitstream remained operational.
In all cases, however, the observed value for $v$ did not change and therefore we set \verb|OCV| to 15.
\begin{proposition}
$\texttt{OCV} \leq N_{\text{threshold}}$ should hold when selecting $N_{\text{threshold}}$ to increase the probability of acknowledged message.
\end{proposition}
\begin{IEEEproof}
Let $m_{i}$ and $m_{i+1}$ be messages with operation frames of length $f_o(m_{i})$ and $f_o(m_{i+1})$, respectively.
When $m_{i}$ is acknowledged, i.e. operation frame of $m_{i}$ has started, the message frame of $m_{i+1}$ starts, causing the remainder of the operation frame of $m_{i}$ to be NACKs in the message frame of $m_{i+1}$.
Since \verb|OCV| acts as upper bound for $f_o(m_{i})$ and $f_o(m_{i+1})$, the message frame of $m_{i+1}$ is now flooded with up to $f_o(m_{i})-1$ NACKs.
$N_{\text{threshold}} < \texttt{OCV}$ would imply that the probability of $m_{i+1}$ getting acknowledged depends on $f_o(m_{i})$ not reaching the upper bound.
\end{IEEEproof}
For all our experiments, we set $N_{\text{threshold}}$ to 20 noting that proper investigation is required to get an optimal value.

\begin{figure}
	\vspace{-2.5mm}
	\subfigure{\includegraphics[width=\columnwidth]{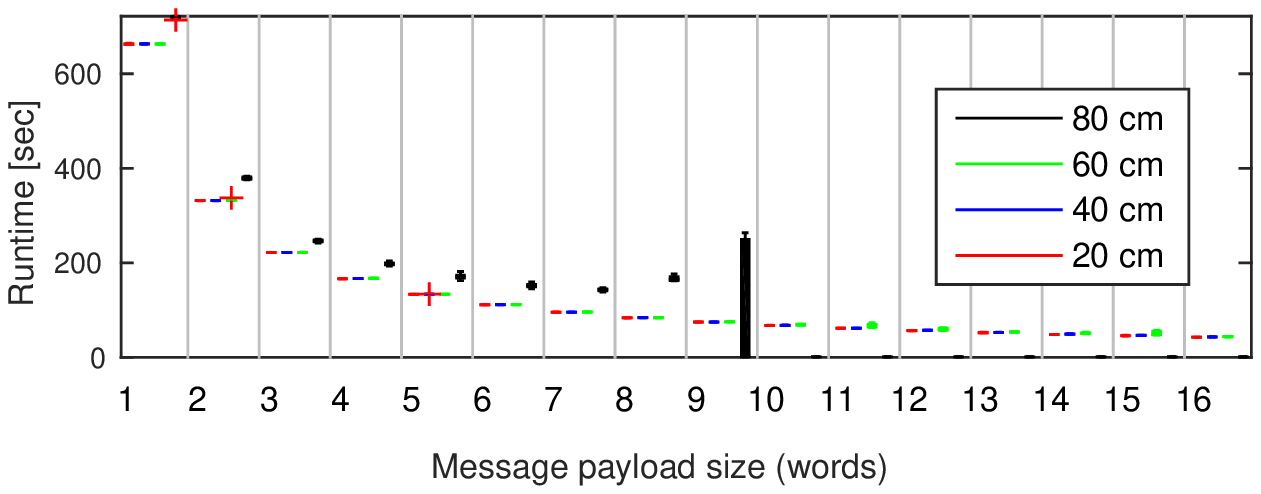}}\vspace{-2.5mm}
	\subfigure{\includegraphics[width=\columnwidth]{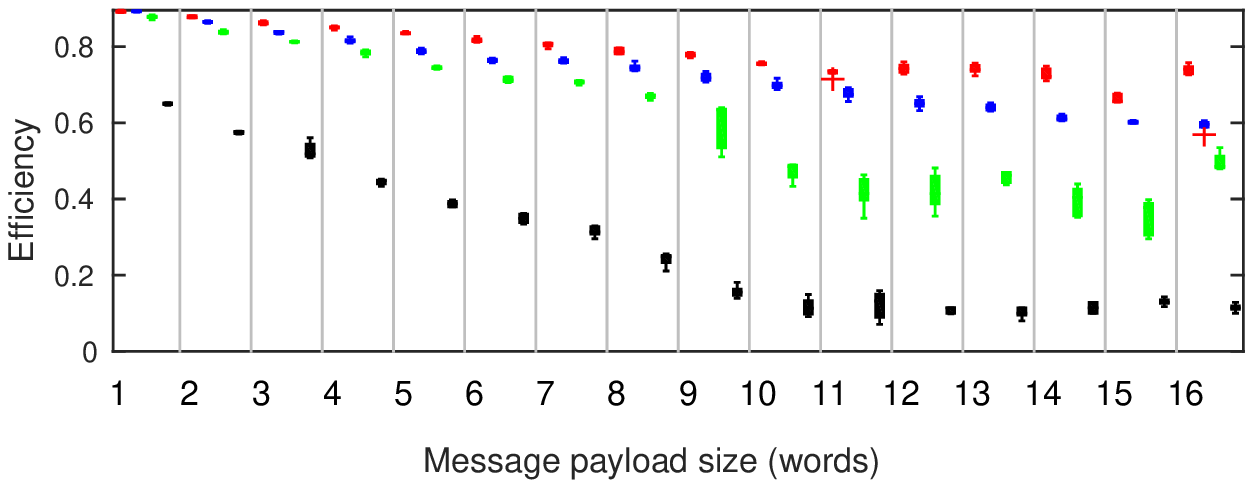}}\vspace{-2.5mm}
	\subfigure{\includegraphics[width=\columnwidth]{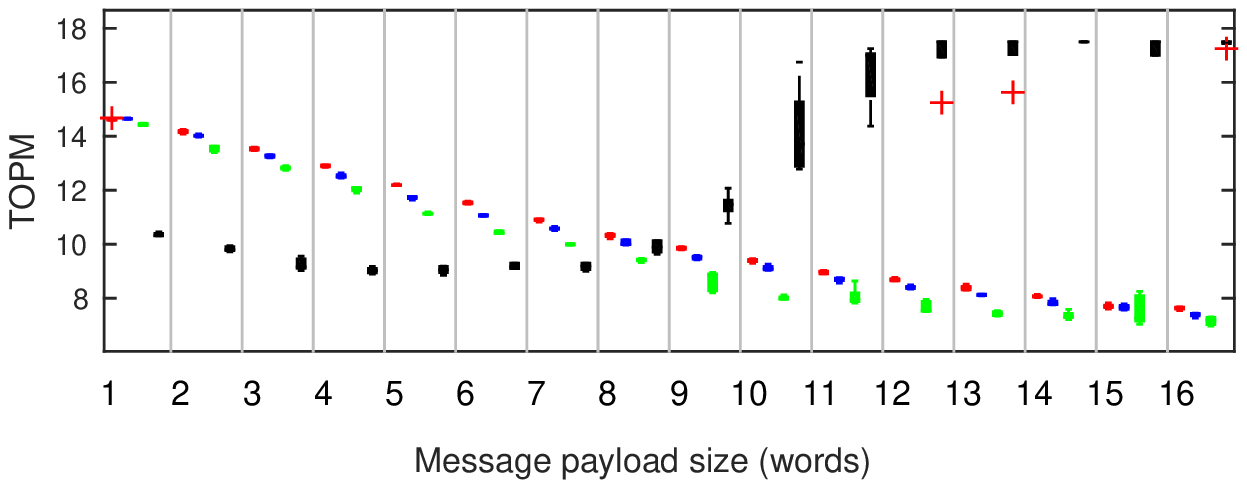}}\vspace{-2.5mm}
	\caption{Runtime, efficiency and total operations per message during a Wisent EX transfer as functions of $S_{\text{p}}$ over various distances.}
	\label{fig:wisentexeval}
\end{figure}

As a first experiment with Wisent EX, now taken in a university \emph{office} instead of laboratory, we transferred Intel Hex files without the throttling mechanism described in~\Cref{sssec:throttling}, containing 5120 bytes of random data, which is around the same amount of bytes as an Intel Hex file of a WISP firmware generated by TI CCS.
However, files generated by TI CCS have a maximum data length of 16\,words per record that cannot be specified by the user.
This forces the record to be split into messages of possibly different $S_{\text{p}}$ values other than the $S_{\text{p}}$ selected for the experiment.
Therefore, we created Intel Hex files for our experiments ourselves in such a way that the records in each of the files hold the same number of words as the value of $S_{\text{p}}$ in the experiment.
The results of each of these experiments with different set $S_{\text{p}}$ over multiple distances $d$ are shown in~\Cref{fig:wisentexeval}.
Assuming a rate of 3.8 messages/sec, the runtime of Wisent Basic would be over two times the amount of Wisent EX using $S_{\text{p}}$ of one word (Wisent Basic uses $S_{\text{p}}=0.5$ and an additional two messages overhead for sending over the address field for each record).
The speedup gained by Wisent EX this way only grows larger, the greater value of $S_{\text{p}}$ is used.
However, the peculiarity in~\Cref{fig:wisentexeval} is the operation of Wisent EX at $d = 60$\,cm, an unstable operating distance in~\Cref{fig:bwreval}.
Instead, this unstable operating distance is shifted to 80\,cm, which can be explained by the presence of more metal objects found in the laboratory than in the office.

\subsubsection{Selection of Values for Throttling Parameters}
\label{sec:optwisentparams}

We now proceed to discuss the procedure of selecting values for throttling parameters in Wisent EX.

For the throttling mechanism, recall that $S_{\text{p}},S_{\text{p+1}} \in T_{t}$ should hold.
Let $T_{t}(S_{\text{p}})$ and $T_{t}(S_{\text{p+1}})$ be the indices of $S_{\text{p}}$ and $S_{\text{p+1}}$ in $T_{t}$.
Now, let $T_{\text{\textsc{u}}}$, $T_{\text{\textsc{de}}}$ and $T_{\text{\textsc{dl}}}$ be denoted as $T_{\text{\textsc{x}}} = T_{t}(S_{\text{p+1}}) - T_{t}(S_{\text{p}})$, i.e. the index difference between $S_{\text{p+1}}$ and $S_{\text{p}}$,
where $T_{\text{\textsc{u}}}$ is used for throttling up, $T_{\text{\textsc{dl}}}$ for throttling down in cases where timeouts are caused by NACKs due the reader losing sight of the CRFID (possibly a more severe case than e.g. NACKs due a mismatched \verb|EPC|), and $T_{\text{\textsc{de}}}$ for throttling down in cases where timeouts are caused by NACKs due any other reason.
We propose the following condition that should hold when selecting values for $T_{\text{\textsc{x}}}$:
\begin{equation}
T_{\text{\textsc{u}}} < \left|T_{\text{\textsc{de}}}\right| \leq \left|T_{\text{\textsc{dl}}}\right|, \qquad |T_{\text{\textsc{x}}}| \in \mathbb{N}^{+}, T_{\text{\textsc{de}}}, T_{\text{\textsc{dl}}} < 0.
\end{equation}
Furthermore, for a message resent for the $R_{\text{max}}$-th time, its $S_{\text{p}}$ should be the minimum possible value, i.e. $S_{\text{p}} = T(1)$, even if before any resend of a message, i.e. $R_{\text{count}} = 0$, $S_{\text{p}}$ was at its maximum possible value, i.e. $S_{\text{p}} = S_{\text{r}}$.
$R_{\text{max}}$ is then found by solving $R_{\text{count}}$ in $\left|T_t\right| - T_{\text{\textsc{de}}} R_{\text{count}} = 1$.

The selected values we have chosen for all system parameters in Wisent EX can be found in~\Cref{tab:paramvalues}.
We conjecture that the value of $M_{\text{threshold}}$ is related to $\left|T_{t}\right|$, since it decides the speed of which $S_{\text{p}}$ converges to a steady state during a set period of time where the communication channel is stable (i.e. the distance from CRFID to antenna is the same for that period).
We note that further analysis should be done to reason about an optimal value for $M_{\text{threshold}}$. We nevertheless feel that a value of 10 supports the rest of values we have chosen for throttling.

\begin{table}
	\centering
	\caption{Selected Values for Wisent EX Throttling Parameters}
	\label{tab:paramvalues}
	\vspace{-2mm}
	\begin{tabular}{|c||c|c|c|c|c|}
		\hline
		Parameter  & $T_{\text{\textsc{u}}}$ & $T_{\text{\textsc{de}}}$ & $T_{\text{\textsc{dl}}}$ & $R_{\text{max}}$ & $M_{\text{threshold}}$    \\ \hline
		Value      & 1                       & --2                       & --3                       & 3                & 10                        \\ \hline
	\end{tabular}
\end{table}

\subsection{Case Study: CRFID Wireless Reprogramming}
\label{sec:evalwireless}

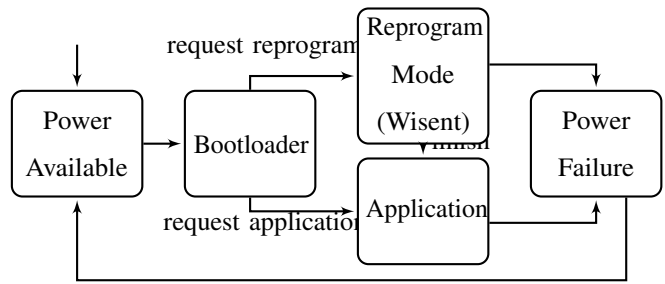
\begin{figure}
	\centering
	\small
	\begin{tikzpicture}[->,>=stealth',shorten >=1pt, semithick, node distance=15mm, scale=1.0, every node/.style={scale=1.0}]
		\draw[use as bounding box, transparent] (-1,-1.8) rectangle (8,1.7);
		\node [thick] at (2.5,1) [above] {request reprogram};
		\node [thick] at (2.5,-1.35) [above] {request application};
		\node [thick] at (5.1,-0.2) [above] {finish};
		
		\node [b] (A) {Power Available};
		\node [b] (B) at ([shift={(2.3,0)}] A)    {Bootloader};
		\node [b] (C) at ([shift={(2.3,0.9)}] B)  {Reprogram Mode (Wisent)};
		\node [b] (D) at ([shift={(2.3,-0.9)}] B) {Application};
		\node [b] (E) at ([shift={(2.3,-0.9)}] C) {Power Failure};
		
		\path [l] (A) --  (B);
		\path [l] (B) |-  (C);
		\path [l] (B) |-  (D);
		\path [l] (C) --  (D);
		\path [l] ([shift={(0,0.4)}]C) -|  (E);
		\path [l] ([shift={(0,-0.4)}]D) -|  (E);
		
		\draw [l] ([shift={(0,0.6)}]A.north) -- (A.north);
		\draw [l] ([shift={(0.4,0)}]E.south) -- ++(0,-1.1) node(lowerright){} |- ([shift={(0,-1.1)}]A.south) -- (A.south);
	\end{tikzpicture}
	\vspace{-2.5mm}
	\caption{Flowchart of the proposed and implemented CRFID bootloader used for wireless reprogramming using Wisent. Compare this design with~\cite[Fig. 3]{ransford2010} and its $n+1$-way switch to determine which firmware program to run.}
	\label{fig:bootloaderabstract}
\end{figure}

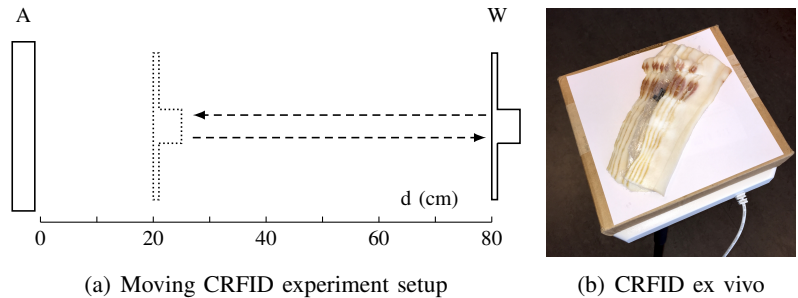
\begin{figure}
	\centering
	\scriptsize
	\subfigure[Moving CRFID experiment setup]{
	\begin{tikzpicture}[semithick, node distance=15mm, scale=1.5]
		\draw[use as bounding box, transparent] (1.9,1) rectangle (6.6,3.3);
		
		\draw[densely dotted, draw=black] (3.3,2) -- (3.5,2) |- (3.5,2.3) -- (3.3,2.3) |- (3.3,2.8) -- (3.25,2.8) |- (3.25, 1.5) -- (3.3,1.5) |- (3.3,2);
		\draw[draw=black, fill=white] (6.3,2) -- (6.5,2) |- (6.5,2.3) -- (6.3,2.3) |- (6.3,2.8) -- (6.25,2.8) |- (6.25, 1.5) -- (6.3,1.5) |- (6.3,2);
		
		\draw [draw=black] (2,1.4) rectangle (2.2,2.9);
		
		\draw [draw, -latex, densely dashed] (3.6,2.05) -- (6.2,2.05);
		\draw [draw, -latex, densely dashed] (6.2,2.25) -- (3.6,2.25);
		
		\draw [draw=black, thin] (2.25,1.3) -- (6.25, 1.3);
		\draw [draw=black, thin] (2.25,1.3) |- (2.25, 1.35);
		\draw [draw=black, thin] (2.75,1.3) |- (2.75, 1.35);
		\draw [draw=black, thin] (3.25,1.3) |- (3.25, 1.35);
		\draw [draw=black, thin] (3.75,1.3) |- (3.75, 1.35);
		\draw [draw=black, thin] (4.25,1.3) |- (4.25, 1.35);
		\draw [draw=black, thin] (4.75,1.3) |- (4.75, 1.35);
		\draw [draw=black, thin] (5.25,1.3) |- (5.25, 1.35);
		\draw [draw=black, thin] (5.75,1.3) |- (5.75, 1.35);
		\draw [draw=black, thin] (6.25,1.3) |- (6.25, 1.35);
		
		\node [thick] at (2.25,1.3) [below] {0};
		\node [thick] at (3.25,1.3) [below] {20};
		\node [thick] at (4.25,1.3) [below] {40};
		\node [thick] at (5.25,1.3) [below] {60};
		\node [thick] at (6.25,1.3) [below] {80};
		\node [thick] at (2.1,3) [above] {A};
		\node [thick] at (6.3,3) [above] {W};
		\node [thick] at (5.7, 1.5) {d (cm)};
	\end{tikzpicture}
	\label{fig:expsetup}
	}
	\subfigure[CRFID ex vivo]{\includegraphics[width=0.2\columnwidth]{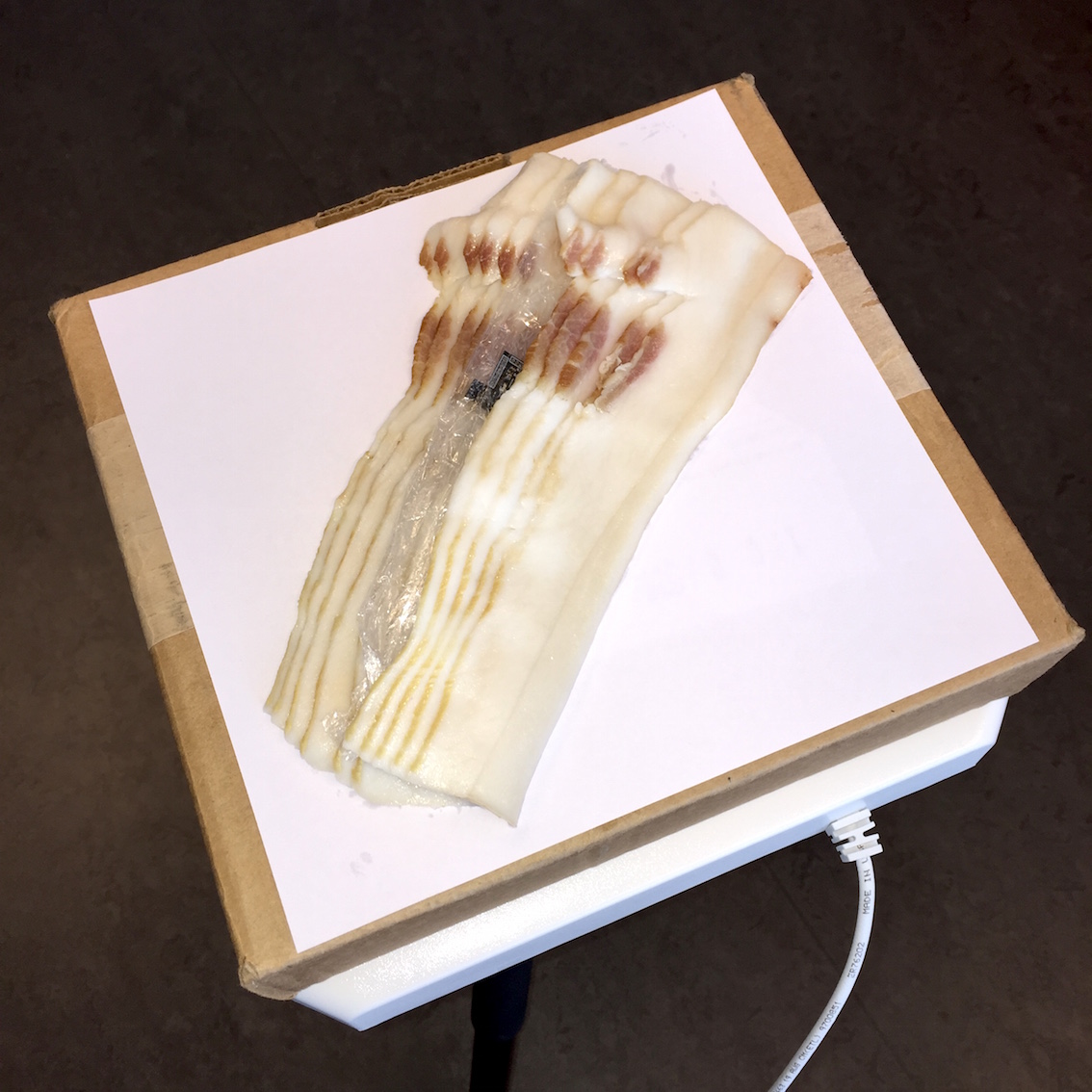}\label{fig:bacon}}
	\caption{Schematic overview of two experiments used to evaluate downstream data transfer with Wisent: (a) moving CRFID (A: antenna, W: WISP), (b) CRFID ex vivo placed between layers of meat (bacon) on RFID antenna.}
\end{figure}

\begin{table}[t]
	\centering
	\begin{threeparttable}
	\caption{WISP Reprogrammed Firmwares}
	\label{tab:firmwarestats}
	\vspace{-2.5mm}
	\begin{tabular}{|l|l|l|}
		\hline
		Firmware                       & Size (bytes) & Runtime\tnote{1} (sec) \\ \hline\hline
		WISP\,5 (base)                 & 5387         &  54.467        \\ \hline
		WISPCam                        & 6442         & 65.097        \\ \hline
		FM0 modulation\tnote{2}        & 1199         &  --- \\ \hline
		Wisent functionality\tnote{2}  & 528          &  --- \\ \hline
	\end{tabular}
	\begin{tablenotes}
		\item[1] \scriptsize {From static distance $d=20$\,cm and with $S_{\text{p}}=16$.}
		\item[2] \scriptsize {Difference of sizes between base firmware with and without this functionality. Runtime omitted as patches were not made.}
	\end{tablenotes}
	\end{threeparttable}
\end{table}

\begin{table}
	\centering
	\begin{threeparttable}
	\caption{Comparison of WISP\,5 Base Firmware Transfer Performance}
	\label{tab:perfcomp}
	\vspace{-1.5mm}
	\begin{tabular}{|r|c|c|c|c|}
		\hline
		$S_{\text{p}}$ & $t$ (sec) & $p_{\text{r}}$ & No. resends & No. transfers completed \\ \hline\hline
		throttle       & 248.448   & 0.0344         & 24.4        & 5/5         \\ \hline
		1              & 810.000   & 0.0004         & 1.20        & 5/5         \\ \hline
		2              & 462.585   & 0.0096         & 13.2        & 5/5         \\ \hline
		3              & 403.974   & 0.0411         & 44.0        & 4/5         \\ \hline
		4              & 351.899   & 0.0951         & 73.0        & 2/5         \\ \hline
		6\tnote{$\star$}     & ---       & ---            & ---         & 0/5         \\ \hline
	\end{tabular}
	\begin{tablenotes}
		\item[$\star$] \scriptsize {$t, p_{\text{r}}$ and no. resends were ommited as all transfers failed due too many resends.}
		\item    \scriptsize {$S_{\text{p}} = \{8,16\}$ were ommited as they show the same result as for $S_{\text{p}}=6$.}
	\end{tablenotes}
	\end{threeparttable}
\end{table}

As final evaluation, we demonstrate the ability to wirelessly reprogram the WISP by using Wisent EX with the selected parameter values in~\Cref{sec:optwisentparams}.
For this, we created a bootloader that is manually programmed to the WISP and is not overwritten after the wireless programming, see~\Cref{fig:bootloaderabstract}.

To initiate the wireless programming, the host sends a transfer initialization message as a \verb|Write| command to enter programming mode.
This command is recognized by the CRFID so that messages are handled correctly until the programming session is finished.
When all data of the application is transferred, the CRC16 checksum over the whole application is sent to validate the firmware.
If the checksum matches the calculated one on the CRFID, the programming session was successful and the bootloader will start the application.

\subsubsection{Experiment Setup and Results}
\label{sssec:reprogramresults}

As experiments, we have adapted a measurement scenario to mimic movement of CRFID and its effect on channel quality using an in-house developed automaton for repeated indoor mobility~\cite{cattani:gondola:2015}.
For this the WISP is attached to nylon wires, which are then wound up/released by stepper motors (controlled by an Arduino) with a speed of approximately 0.1\,m/s to move the WISP and change its distance to the antenna between 20\,cm and 90\,cm (i.e. between optimal distance and a distance where channel is unreliable) in a repeatable manner, see~\Cref{fig:expsetup}.
Results of these experiments using the constructed bootloader are listed in~\Cref{tab:firmwarestats} and~\Cref{tab:perfcomp}, which justifies the use of the throttling mechanism as proven in~\Cref{sssec:throttling}.
In comparison with the case of a set $S_{\text{p}}=4$ the throttling mechanism cuts the transfer time of the base firmware down by approximately 100\,seconds and even reduces the resend rate with almost three times while finishing all transfer attempts.

\subsubsection{Data Transfer Energy Consumption}
\label{sssec:energy_consumption}

For result completeness and comparison to typical state-of-the-art WSN devices concerning the energy consumption used in downstream communication, we have used a Monsoon Power Monitor to measure the energy consumption of a Tmote Sky node storing 5387\,bytes of data received from another node (WISP\,5 base firmware length, see~\Cref{tab:firmwarestats}).
To represent Wisent EX messages in our experiment, the data was sent in chunks of 36 bytes with the X-MAC protocol present in Contiki OS.
The measured energy consumption of the Tmote Sky is 256.35\,mJ, while the WISP consumed a total of 81.70\,mJ.
This proves that despite Wisent not being designed with energy efficiency in mind, CRFID downstream transfer is at least three times more energy efficient than corresponding WSN downstream.

\subsubsection{Reprogramming Tissue-Embedded CRFID}
\label{sec:implantsim}

As an ultimate experiment, we demonstrate the ability to wirelessly reprogram a tissue-implanted CRFID. To emulate such a scenario we placed a cling film-wrapped WISP between 5 and 6 layers of meat (bacon) at the top and bottom of the WISP, respectively, placed on an antenna separated by a 6\,cm paper box, see Fig.~\ref{fig:bacon}. This experiment followed similar ex vivo experiments emulating implantable sensor scenarios~\cite[Fig. 8]{halperin:sap:2008}.
We were able to successfully reprogram the WISP with complete RFID stack within 63.55\,sec despite attenuation of the backscatter signals from the meat.


\section{Limitations and Future Work}
\label{sec:futurework}

Wisent forms a baseline for experiments on downstream CRFID communication. Further required features are:
\begin{enumerate}
\item \emph{Transfer to multiple tags at the same time}: although Wisent has been designed for transfers to a single CRFID tag, an extension for multiple tags is necessary which involves careful scheduling of resources;
\item \emph{Security}: to deploy reprogrammable CRFIDs, data transfer needs to be secured. Wisent as of now has no mechanism to prevent message spoofing. We argue that this is the most urgent feature missing in Wisent. 
\end{enumerate}
Other functionalities requiring consideration include the ability to resume a transfer after failure or addition of a data compression mechanism, which has a trade-off of performance versus computation power used by the CRFID.


\section{Conclusion}
\label{sec:conclusion}

In this paper, we have designed and implemented a protocol, called Wisent, that allows to transfer bulk data from host to CRFID in a fast and robust manner. Wisent allows to store and verify data despite power interruptions at the CRFID thanks to the use of non-volatile FRAM memory and simple error verification mechanism. In addition, through introduction of large frame sizes (sent by the RFID reader), thanks to such ability of EPC C1G2 RFID communication protocol, and its length adaptations depending on the channel conditions, Wisent improves the throughput threefold in comparison to single word message size supported by the EPC C1G2 standard. Finally, implementation of Wisent allowed to introduce and experimentally verify the world's first wirelessly reprogrammable (software defined) CRFID.


\end{document}